\documentclass[aps,pre,groupedaddress,10pt]{revtex4-1}
\usepackage[active]{srcltx}
\usepackage{graphicx}
\usepackage{amsmath, amssymb, amsfonts}
\usepackage{color}

\begin{document}

%\preprint{}

\title{Weakening connections in heterogeneous mean-field models}


\author{C. Dias$^{1}$ and M. O. Hase$^{1}$}
%\thanks{}
\affiliation{$^{1}$ Escola de Artes, Ci\^encias e Humanidades, Universidade de S\~ao Paulo, Av. Arlindo B\'ettio 1000, 03828-000 S\~ao Paulo, Brazil}


%\date{\today}

\begin{abstract}
Two versions of the susceptible-infected-susceptible epidemic model, which have different transmission rules, are analysed. Both models are considered on a weighted network to simulate a mitigation in the connection between the individuals. The analysis is performed through a heterogeneous mean-field approach on a scale-free network. For a suitable choice of the parameters, both models exhibit a positive infection threshold, when they share the same critical exponents associated with the behaviour of the prevalence against the infection rate. Nevertheless, when the infection threshold vanishes, the prevalence of these models display different algebraic decays to zero for low values of the infection rate.
\end{abstract}

\email{mhase@usp.br}

%89.75.Hc	Networks and genealogical trees
%02.50.Ga	Markov processes
%02.50.Ey	Stochastic processes

%\keywords{}

\maketitle


 %%%%%%%%%%%%%%%%%%%%%%%%%%%%%%%%%%%%%%%%%%%%%%%%%%%%%%%%%%%%%%%%%%%%%%%%%%%%%%%%%%%%%%%%%%%%%%%%%%%%
 %%%%%%%%%%%%%%%%%%%%%%%%%%%%%%%%%%%%%%%%%%%%%%%%%%%%%%%%%%%%%%%%%%%%%%%%%%%%%%%%%%%%%%%%%%%%%%%%%%%%
 %%%%%%%%%%%%%%%%%%%%%%%%%%%%%%%%%%%%%%%%%%%%%%%%%%%%%%%%%%%%%%%%%%%%%%%%%%%%%%%%%%%%%%%%%%%%%%%%%%%%

\section{Introduction}

Recently, the COVID-19 outbreak imposed a new lifestyle to face this disease. It also reminded us that new infectious diseases can emerge and the difficulties to make predictions on the course of its diffusion based on incomplete data (for instance, due to underreporting) have emphasized the importance of mathematical modelling of infection spreading \cite{H89, AM92}.

A simple and standard scheme that is popular in mathematical modelling in epidemiology is the division of the population into compartments, which characterises the possible states of individuals with respect to the disease: infected, susceptible, recovered, \textit{et c\ae tera}. The main strategy is to write a set of differential equations that describes the flow of the population between the compartments. Traditionally, these equations have adopted the hypothesis of \textquotedblleft homogeneous mixing\textquotedblright, in which an individual has the same probability of being in contact with any member of the compartment \cite{H06}. Although this supposition has been popular for decades, it leads to an oversimplification that contrasts with the heterogeneous organization in the network of human contacts \cite{AB02, DM03, N10}. This observation has pointed out the importance of combining both epidemiology and network theory in order to study the spreading of infectious agents based on a realistic interconnection between individuals \cite{PSCvMV15}.

Apart from the complex structure of the human contact network that is vital to monitor the spreading of diseases, the recent COVID-19 pandemic has reinforced the importance of keeping social distance, especially when no other efficient alternative to contain the propagation of the infection is known. As a consequence, the physical ties between people are effectively weakened, either by deliberate seclusion or by forced isolations (like hospitalisations). In this work, we examine the effects of the epidemic spreading in a scenario where the contact between individuals is mitigated (especially, the contact to infected ones). The main strategy to accomplish this goal is to introduce a tunable parameter $\omega$ that controls the probability of linking between the members of the system. This modification is equivalent to analyse a class of weighted networks \cite{BBPSV04}, and the definition of epidemiological models on such graphs has also been considered by several authors \cite{CGZZ09, CZGZ10}.

We test our ideas on the susceptible-infected-susceptible (SIS) model, where each member of the system (represented by the vertices of a network) can be in one of two states: susceptible or infected. The former can be infected by direct contact with an infected individual, while the latter can recover spontaneously to integrate the set of susceptible members. We perform the analysis of the SIS dynamics, with weakening effects in the connections between individuals, on two similar - but different - systems. In the investigation of infection spreading, many transmission mechanisms can be proposed \cite{M16}. We highlighted two of them, which we call, following \cite{CMF18}, the \textit{standard} SIS (s-SIS), and \textit{threshold} SIS (t-SIS) models. In the former, an infected node infects each susceptible neighbour with a rate $\lambda$. On the other hand, in t-SIS model, a susceptible node is infected, with rate $\lambda$, if at least one of its neighbours is infected, thus characterising a threshold process \cite{DB88, BHH18} (see also \cite{D88, E89}, where a similar idea is present in the \textquotedblleft first epidemic model\textquotedblright). Our investigation for both s-SIS and t-SIS models is based on the heterogeneous mean-field (HMF) approximation \cite{PSV01a, PSV01b}, which allows the inclusion of heterogeneities (present in real networks) at the level of degree distributions. Despite this appealing feature of the HMF, it allows a finite infection threshold for a SIS dynamics on a scale-free network \cite{PSV01b}. In other words, it implies the existence of a critical point that separates a disease-free absorbing state from the active phase, where the infection persists. This is in contrast with a rigorous result that states the location of this critical point to be at zero for an uncorrelated scale-free random network \cite{CD09}. The vanishment of the critical point is, however, predicted by the SIS model in the quenched mean-field (QMF) \cite{HY84, WCWF03} approach for networks that are governed by a power-law degree distribution \cite{CPS10, BCPS13}. The QMF captures the structure of the network through its adjacency matrix, and the discrepancy coming from the quenched or annealed (of which HMF is an example) treatment in many statistical models is not surprising \cite{HM08, CH16}. All the models mentioned above are usually treated at mean-field level, whose analysis does not mandatorily demand metric features of the problem. Nevertheless, the incorporation of the information based on spatial structure of the environment, like in metapopulation models \cite{CPSV07, CV08, MFPS13}, is beneficial to epidemiology.

The layout of this paper is as follows. We briefly review the s-SIS and t-SIS models in Section 2 to introduce some notations. In Section 3, we show the modification necessary in these models to introduce weakening effects on the links between individuals. Then, we examine the s-SIS and t-SIS models with weakening in connections in Sections 4 and 5, respectively. The results from both models are compared in Section 6 and the conclusions are delivered in the last Section.

 %%%%%%%%%%%%%%%%%%%%%%%%%%%%%%%%%%%%%%%%%%%%%%%%%%%%%%%%%%%%%%%%%%%%%%%%%%%%%%%%%%%%%%%%%%%%%%%%%%%%
 %%%%%%%%%%%%%%%%%%%%%%%%%%%%%%%%%%%%%%%%%%%%%%%%%%%%%%%%%%%%%%%%%%%%%%%%%%%%%%%%%%%%%%%%%%%%%%%%%%%%
 %%%%%%%%%%%%%%%%%%%%%%%%%%%%%%%%%%%%%%%%%%%%%%%%%%%%%%%%%%%%%%%%%%%%%%%%%%%%%%%%%%%%%%%%%%%%%%%%%%%%

\section{Heterogeneous mean-field models}
\label{hmf}

In this work, we examine the SIS model, where an infected individual infects, by contact, a susceptible one with a rate $\lambda$, and an infected recovers spontaneously with a rate $\mu$. In an uncorrelated network, the assumption of the \textquotedblleft homogeneous mixing\textquotedblright hypothesis leads to an infection-free absorbent phase for a sufficiently low infection rate. Nevertheless, increasing this rate over a critical threshold $\lambda_{c}$, which is inversely proportional to the mean degree of the graph, implies the persistence of the disease \cite{BPSV03}.

Henceforth, we consider the cases where the network of contacts does not have a homogeneous structure. An epidemiological model defined on a heterogeneous (in the sense of degree distribution) network was first investigated in the framework of a HMF theory \cite{PSV01a, PSV01b}. In this approach, the time evolution of $\rho_{k}$, which is the probability of a vertex with degree $k$ being infected, is given by
\begin{align}
\partial_{t}\rho_{k} = -\rho_{k} + \lambda k\left(1-\rho_{k}\right)\Theta_{k},
\label{sSIS}
\end{align}
where $\partial_{t}\equiv\frac{\partial}{\partial t}$ stands for the (partial) derivative with respect to time. This is a s-SIS model, where an infected vertex infects each susceptible neighbour with a rate $\lambda$. We treat vertices that share the same degree on equal footing, but we will see that the probability of being infected differs if they have different connectivity. In \eqref{sSIS}, the recovering rate $\mu$ is taken to be $\mu=1$ without loss of generality. This procedure is always possible by choosing a suitable time scale, and then the infection rate $\lambda$ becomes numerically equivalent to the spreading rate $\lambda/\mu$. In the last term, $\Theta_{k}$ stands for the probability that a link originated from a vertex of degree $k$ connects to an infected node.

In this work, we consider an uncorrelated network, where the probability $\Theta_{k}$ is independent of $k$, and we replace this symbol by $\Theta$. As a consequence of this assumption, the probability $q(k|k^{\prime})$ of a vertex of degree $k^{\prime}$ linking to a node of degree $k$ can be cast as \cite{N02, DM03}
\begin{align}
q(k|k^{\prime}) = q(k) = \frac{kP(k)}{\langle k\rangle},
\label{q}
\end{align}
where $P$ is the degree distribution and $\langle k^{\alpha}\rangle:=\sum_{k}k^{\alpha}P(k)$ stands for the $\alpha$-th moment. For simplicity, we write $q(k)$ instead of $q(k|k^{\prime})$, since $q$ does not depend on the degree of the source vertex in an uncorrelated network. Therefore, the probability $\Theta$ can be written as
\begin{align}
\Theta = \sum_{k}q(k)\rho_{k} = \frac{1}{\langle k\rangle}\sum_{k}kP(k)\rho_{k},
\label{Theta}
\end{align}
which is the sum of all disjoint events of linking to an infected node of degree $k$.

Contrary to the s-SIS model described above, we may conceive a different infection mechanism, where a susceptible vertex changes its state if at least one of its neighbours is infected \cite{M16}. This is the t-SIS model, since the infection is a threshold process. The HMF master equation in this scenario is given by
\begin{align}
\partial_{t}\rho_{k} = -\rho_{k} + \lambda\left(1-\rho_{k}\right)\left[1-\left(1-\Theta\right)^{k}\right].
\label{tSIS}
\end{align}
Again, a suitable choice in time scaling leads to a unitary recovery rate. As in the previous model, we consider an uncorrelated network, and the factor $\left(1-\Theta\right)^{k}$ is the probability that none of the $k$ neighbours of a vertex is infected, which follows from the assumption that it constitutes $k$ independent events. As a consequence, $1-\left(1-\Theta\right)^{k}$ is the probability that at least one of the $k$ neighbours is infected. Despite some similarities (shown later), this is a different model from the s-SIS, although some confusion may arise \cite{M16}.

In this work, we mainly investigate the stationary state, where $\partial_{t}\rho_{k}=0$. The quantification of the infection is examined through the prevalence
\begin{align}
\rho = \sum_{k}P(k)\rho_{k},
\label{prevalence}
\end{align}
which is the stationary infection probability of the system.

 %%%%%%%%%%%%%%%%%%%%%%%%%%%%%%%%%%%%%%%%%%%%%%%%%%%%%%%%%%%%%%%%%%%%%%%%%%%%%%%%%%%%%%%%%%%%%%%%%%%%
 %%%%%%%%%%%%%%%%%%%%%%%%%%%%%%%%%%%%%%%%%%%%%%%%%%%%%%%%%%%%%%%%%%%%%%%%%%%%%%%%%%%%%%%%%%%%%%%%%%%%
 %%%%%%%%%%%%%%%%%%%%%%%%%%%%%%%%%%%%%%%%%%%%%%%%%%%%%%%%%%%%%%%%%%%%%%%%%%%%%%%%%%%%%%%%%%%%%%%%%%%%

\section{SIS models with weakened connections}
\label{weak}

In this section, a modification in the models presented in the previous section is proposed, where the connections are weakened. We mean by this that the probability $\Theta$ of linking to an infected vertex is decreased. This situation may simulate a scenario where infected individuals become harder to be reached (by isolating themselves, being hospitalised, \textit{et c\ae tera}). The modification is introduced by replacing the linking probability \eqref{q} by
\begin{align}
q_{\omega}(k) = \frac{k^{\omega}P(k)}{\langle k^{\omega}\rangle},
\label{qw}
\end{align}
where $\omega$ is a parameter that we assume to belong in the interval $(-\infty,1]$. The case $\omega=1$ recovers the usual case on an uncorrelated network. Therefore, the probability of a link achieving an infected link is
\begin{align}
\Theta_{\omega} = \sum_{k}q_{\omega}(k)\rho_{k} = \frac{1}{\langle k^{\omega}\rangle}\sum_{k}k^{\omega}P(k)\rho_{k}.
\label{Thetaw}
\end{align}
The index $\omega$ in $\Theta_{\omega}$ is written to remember that we are dealing with the modified version of the model.

It is important to stress that the probabilities $q$ and $q_{\omega}$ are quantities associated to graph properties and are not related to the epidemiological context. Nevertheless, they appear, respectively, in the probabilities $\Theta$ and $\Theta_{\omega}$ only. This is the reason we can treat the replacement of $\Theta$ by $\Theta_{\omega}$ as a weakening of the links to infected vertices, because contacts between susceptible ones do not play any role in SIS dynamics. The model defined in \eqref{Thetaw} is equivalent to define the SIS model on weighted networks \cite{BBPSV04}.

 %%%%%%%%%%%%%%%%%%%%%%%%%%%%%%%%%%%%%%%%%%%%%%%%%%%%%%%%%%%%%%%%%%%%%%%%%%%%%%%%%%%%%%%%%%%%%%%%%%%%
 %%%%%%%%%%%%%%%%%%%%%%%%%%%%%%%%%%%%%%%%%%%%%%%%%%%%%%%%%%%%%%%%%%%%%%%%%%%%%%%%%%%%%%%%%%%%%%%%%%%%
 %%%%%%%%%%%%%%%%%%%%%%%%%%%%%%%%%%%%%%%%%%%%%%%%%%%%%%%%%%%%%%%%%%%%%%%%%%%%%%%%%%%%%%%%%%%%%%%%%%%%

\section{Standard SIS model with weakened connections}
\label{sSISw}

\subsection{General setting}

\begin{figure}
\begin{center}
\includegraphics[width=200pt]{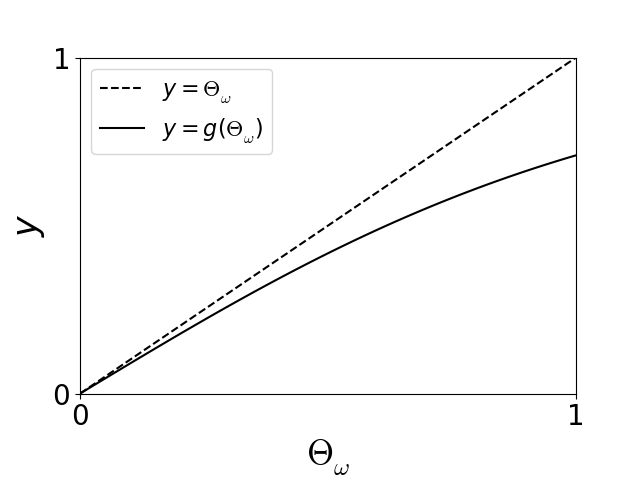}
\includegraphics[width=200pt]{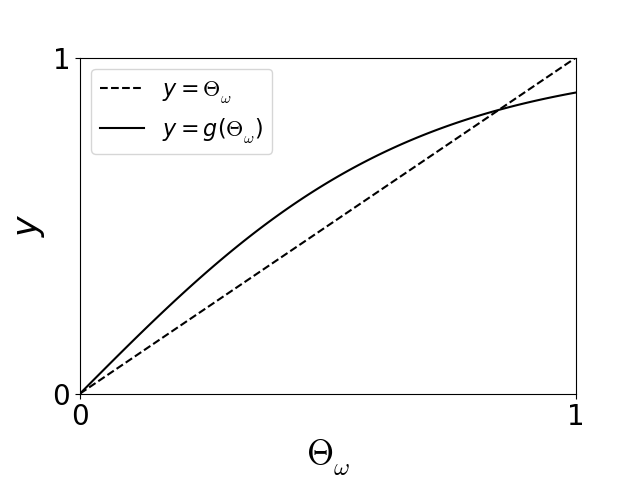}
\caption{\label{tgtfig} Sketch of the graphical solution of \eqref{TgT}. Left: Trivial solution at $\Theta_{\omega}=0$. Right: Trivial solution at $\Theta_{\omega}=0$ and a non-trivial positive solution.}
\end{center}
\end{figure}

We examine the stationary state of the s-SIS model, which consists of the master equation \eqref{sSIS} and the probability \eqref{Thetaw}. Some of the procedures are well-known, but we present them for completeness. In steady state, $\partial_{t}\rho_{k}=0$, equation \eqref{sSIS} implies
\begin{align}
\rho_{k} = \frac{\lambda k\Theta_{\omega}}{1+\lambda k\Theta_{\omega}},
\label{ssis_rhok}
\end{align}
which shows the inhomogeneous structure of the network, because the infection probability depends on the degree of the vertex. Inserting \eqref{ssis_rhok} into \eqref{Thetaw} leads to
\begin{align}
\Theta_{\omega} = g(\Theta_{\omega}),
\label{TgT}
\end{align}
where
\begin{align}
g(\Theta_{\omega}) = \frac{1}{\langle k^{\omega}\rangle}\sum_{k}k^{\omega}P(k)\frac{\lambda k\Theta_{\omega}}{1+\lambda k\Theta_{\omega}}.
\label{g}
\end{align}
Since $\Theta_{\omega}$ is a probability, one has $0\leq\Theta_{\omega}=g(\Theta_{\omega})\leq 1$. It is also immediate that $g(0)=0$, $g^{\prime}(\Theta_{\omega})>0$ and $g^{\prime\prime}(\Theta)<0$ (the symbol $g^{\prime}$ stands for the derivative of $g$). Hence, $g$ is an increasing and concave function in $[0,1]$ such that $g(0)=0$ and $g(1)\leq 1$. This last upper bound follows also from
\begin{align}
\nonumber g(\Theta_{\omega}=1)&=\frac{1}{\langle k^{\omega}\rangle}\sum_{k}k^{\omega}P(k)\frac{\lambda k\Theta_{\omega}}{1+\lambda k\Theta_{\omega}} \\
  &\leq\frac{1}{\langle k^{\omega}\rangle}\sum_{k}k^{\omega}P(k)\frac{1+\lambda k\Theta_{\omega}}{1+\lambda k\Theta_{\omega}}=1.
\label{upperbound}
\end{align}
The equation \eqref{TgT} can be solved by searching for the intersection points of the graphs $y=\Theta_{\omega}$ and $y=g(\Theta_{\omega})$ in the plane $y\times\Theta_{\omega}$, as one can see in figure \ref{tgtfig}. The critical point is obtained from the condition $g^{\prime}(0)=1$, which implies
\begin{align}
\lambda_{c} = \frac{\langle k^{\omega}\rangle}{\langle k^{\omega+1}\rangle}.
\label{lambdawc}
\end{align}
This result could have been obtained through a simpler approach \cite{M16}. However, by the present method we can be sure that there are at most two solutions in the model.

The analysis of the system above is based on the equations \eqref{TgT} and \eqref{g} with the constraint $\Theta_{\omega}\neq 0$. Concretely, it implies the investigation of the equation
\begin{align}
\frac{1}{\langle k^{\omega}\rangle}\int_{0}^{\infty}\textup{d}k\,k^{\omega}P(k)\frac{\lambda k}{1+\lambda k\Theta_{\omega}} = 1,
\label{principal_ssis}
\end{align}
where the continuum approximation was invoked.

 %%%%%%%%%%%%%%%%%%%%%%%%%%%%%%%%%%%%%%%%%%%%%%%%%%%%%%%%%%%%%%%%%%%%%%%%%%%%%%%%%%%%%%%%%%%%%%%%%%%%

\subsection{Scale-free network}
\label{caseSF}

We examine the s-SIS model on a scale-free network, where the degree distribution scales as $P(k)\sim k^{-\gamma}$, with weakened connections gauged by the parameter $\omega$. Let us first establish some notations and results. Assuming that the minimum degree of each vertex in the network is $m$, one has
\begin{align}
P(k) = \frac{\left(\gamma-1\right)m^{\gamma-1}}{k^{\gamma}} \quad (k\geq m).
\label{SF}
\end{align}
We assume that $\gamma>2$ to ensure the existence of
\begin{align}
\langle k^{\omega}\rangle=\frac{\left(\gamma-1\right)m^{\omega}}{\gamma-\omega-1} \quad (0\leq\omega\leq 1),
\label{<kw>}
\end{align}
where the average is taken with respect to the distribution \eqref{SF}. From \eqref{lambdawc}, the critical point is
\begin{align}
\lambda_{c} = \frac{1}{m}\left(\frac{\gamma-\omega-2}{\gamma-\omega-1}\right),
\label{lambdawcSF}
\end{align}
and we have to admit the condition $\gamma-\omega-2>0$ to ensure a positive infection threshold.

As seen before, the s-SIS model above the infection threshold is described by \eqref{principal_ssis}. On the scale-free network \eqref{SF}, this equation is cast as
\begin{align}
\left(\gamma-\omega-1\right)\lambda m\epsilon^{\gamma-\omega-2}\int_{\epsilon}^{\infty}\frac{\textup{d}u}{u^{\gamma-\omega-1}\left(1+u\right)} = 1,
\label{principalSF}
\end{align}
where $\epsilon:=\lambda m\Theta_{\omega}$. We are especially interested in the behaviour of the system close to the infection threshold. In this region, the integral in \eqref{principalSF} can be examined analytically, as shown in Appendix \ref{apA}. Here, we will just state the results.

The evaluation of the prevalence \eqref{prevalence} in s-SIS model close to the critical point benefits also from the results of Appendix \ref{apA}. When $\gamma>2$, we have
\begin{align}
\rho\simeq\left(\frac{\gamma-1}{\gamma-2}\right)\left(\lambda m\Theta_{\omega}\right),
\label{rhossisafter}
\end{align}
where
\begin{align}
\lambda m\Theta_{\omega}\simeq\left\{
\begin{array}{lcl}
\displaystyle\left[\frac{m\left(\gamma-\omega-1\right)\pi}{\sin\left((\gamma-\omega-1)\pi\right)}\right]^{\frac{1}{2-\gamma+\omega}}\lambda^{\frac{1}{2-\gamma+\omega}} & , & 1<\gamma-\omega<2 \\
 & & \\
\displaystyle\left[\frac{\sin\left(-(\gamma-\omega-1)\pi\right)}{\lambda\pi\left(\gamma-\omega-2\right)}\right]^{\frac{1}{\gamma-\omega-2}}\left(\lambda-\lambda_{c}\right)^{\frac{1}{\gamma-\omega-2}} & , & 2<\gamma-\omega<3 \\
 & & \\
\displaystyle\frac{1}{\lambda}\left(\frac{\gamma-\omega-3}{\gamma-\omega-2}\right)\left(\lambda-\lambda_{c}\right) & , & \gamma-\omega>3
\end{array}
\right..
\label{Thetassis}
\end{align}
This result recovers the ones obtained by \cite{PSV01b}.

The following conclusions can be established. Firstly, the infection threshold vanishes if $\gamma-\omega\leq 2$. However, the way the infection probability $\rho$ (and also $\Theta_{\omega}$) approaches zero should be examined carefully. For $\gamma-\omega<2$, we observe an algebraic decay $\rho\sim\lambda^{\frac{1}{2-\gamma+\omega}}$, while a non-algebraic decay, $\rho\sim e^{-\frac{1}{\lambda m}}$, is predicted if $\gamma-\omega=2$. This last case is realized \cite{PSV01a} in the usual s-SIS model defined on a Barab\'asi-Albert network \cite{BA99}.

For $\gamma-\omega>2$, a positive critical point emerges. In the interval $2<\gamma-\omega<3$, we see an algebraic decay of the order parameter $\rho\sim\left(\lambda-\lambda_{c}\right)^{\frac{1}{\gamma-\omega-2}}$. On the other hand, if $\gamma-\omega=3$, the relation $\Theta_{\omega}\ln\Theta_{\omega}\sim\left(\lambda-\lambda_{c}\right)$ holds close to the critical point, discarding an algebraic relation between $\Theta_{\omega}$ (and, consequently, $\rho$) and the difference $\lambda-\lambda_{c}$. Finally, if $\gamma-\omega>3$, the criticality of the system is described by $\rho\sim\left(\lambda-\lambda_{c}\right)$.

From \eqref{Thetassis}, the probability $\Theta_{\omega}$ depends on the parameters $\gamma$ and $\omega$ through its difference $\gamma-\omega$, and this property is observed in both critical exponents and the prefactors. However, this type of dependence is broken when we consider the prevalence $\rho$ due to the factor $\frac{\gamma-1}{\gamma-2}$ in \eqref{rhossisafter}.

\begin{figure}
\begin{center}
\includegraphics[width=243pt]{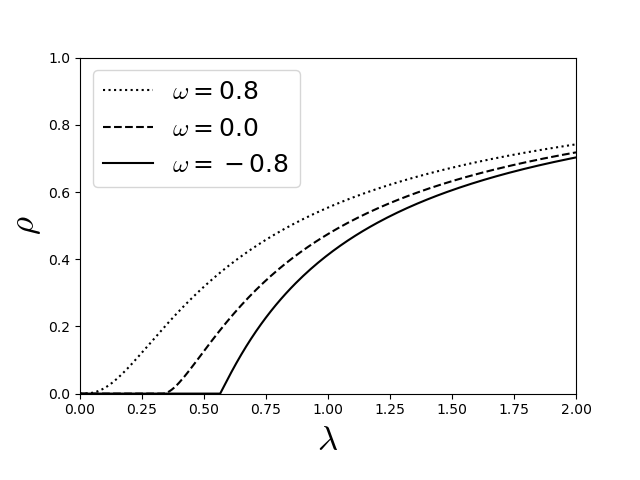}
\caption{\label{rhoxlambdaSF} Graph $\rho\times\lambda$ for the s-SIS model with weakened connections; here, $\gamma=2.5$. The curves were obtained from numerical resolution of equation \eqref{principalSF}. Although the prevalence does not depend on $\lambda$ and $m$ independently (actually, it is a function of $\lambda m$, as one can easily see from \eqref{principalSF}), we fixed the value of $m$ to $1$ to show the dependence on the infection rate only.}
\end{center}
\end{figure}

 %%%%%%%%%%%%%%%%%%%%%%%%%%%%%%%%%%%%%%%%%%%%%%%%%%%%%%%%%%%%%%%%%%%%%%%%%%%%%%%%%%%%%%%%%%%%%%%%%%%%
 %%%%%%%%%%%%%%%%%%%%%%%%%%%%%%%%%%%%%%%%%%%%%%%%%%%%%%%%%%%%%%%%%%%%%%%%%%%%%%%%%%%%%%%%%%%%%%%%%%%%
 %%%%%%%%%%%%%%%%%%%%%%%%%%%%%%%%%%%%%%%%%%%%%%%%%%%%%%%%%%%%%%%%%%%%%%%%%%%%%%%%%%%%%%%%%%%%%%%%%%%%

\section{Threshold SIS model with weakened connections}
\label{tSISw}

In this section, we investigate the stationary state of the t-SIS model defined by the master equation \eqref{tSIS}. In this regime, one has
\begin{align}
\rho_{k} = \frac{\lambda\left[1-\left(1-\Theta\right)^{k}\right]}{1+\lambda\left[1-\left(1-\Theta\right)^{k}\right]} = 1 - \frac{1}{1+\lambda\left[1-\left(1-\Theta\right)^{k}\right]}.
\label{rhoTT}
\end{align}
Let us now include the weakening factor in the model by inserting \eqref{rhoTT} into \eqref{Thetaw}, which leads to
\begin{align}
\Theta_{\omega} = 1 - \frac{1}{\langle k^{\omega}\rangle}\sum_{k}k^{\omega}P(k)\frac{1}{1+\lambda\left[1-\left(1-\Theta_{\omega}\right)^{k}\right]},
\label{ThetaT}
\end{align}
where the degree distribution is given by \eqref{SF} and the mean value $\langle\cdot\rangle$ is taken with respect to this scale-free distribution. We assume, again, that the minimum degree of the graph is $m$. Define, now, the probability
\begin{align}
\varphi_{\omega} := 1 - \Theta_{\omega},
\label{varphi}
\end{align}
which is just the complementary one to $\Theta_{\omega}$. From this notation, our principal equation, \eqref{ThetaT}, can be cast as
\begin{align}
\varphi_{\omega} = h(\varphi_{\omega}),
\label{varphih}
\end{align}
where
\begin{align}
h(\varphi_{\omega}) := \frac{1}{\langle k^{\omega}\rangle}\sum_{k}k^{\omega}P(k)\frac{1}{1+\lambda\left(1-\varphi_{\omega}^{k}\right)}.
\label{h}
\end{align}
Noting from \eqref{h} that $h(0)\leq 1$, $h(1)=1$, $h^{\prime}(\varphi_{\omega})>0$ and $h^{\prime\prime}(\varphi_{\omega})>0$, and following a similar reasoning of section \ref{sSISw} that led to \eqref{lambdawc}, we can determine the critical point $\lambda_{c}$ through the relation $h^{\prime}(1)=1$, which implies
\begin{align}
\lambda_{c} = \frac{\langle k^{\omega}\rangle}{\langle k^{\omega+1}\rangle},
\label{lambdawcT}
\end{align}
and is exactly the same value obtained in the previous s-SIS model \cite{M16}.

The system \eqref{varphih}-\eqref{h} has a trivial solution $\varphi_{\omega}=1$, which corresponds to $\Theta_{\omega}=0$. We will now search for the nontrivial solutions of this system, and keep in mind that $\varphi\neq 1$. In continuum approximation, we have
\begin{align}
\nonumber\varphi_{\omega} &= \frac{1}{\langle k^{\omega}\rangle}\int_{0}^{\infty}\textup{d}k\,k^{\omega}P(k)\frac{1}{1+\lambda\left(1-\varphi_{\omega}^{k}\right)} \\
 &= \left(\gamma-\omega-1\right)m^{\gamma-\omega-1}\int_{m}^{\infty}\frac{\textup{d}k}{k^{\gamma-\omega}}\frac{1}{1+\lambda\left(1-\varphi_{\omega}^{k}\right)}.
\label{principalT}
\end{align}
The asymptotic behaviour for $\varphi\sim 1^{-}$ of the integral in \eqref{principalT} is examined with detail in Appendix \ref{apB}. From this result, we can characterise the prevalence \eqref{prevalence} in this model, which is
\begin{align}
\rho\simeq\left(\frac{\gamma-1}{\gamma-2}\right)\left(\lambda m\Theta_{\omega}\right),
\label{rhotsisafter}
\end{align}
where
\begin{align}
\lambda m\Theta_{\omega}\simeq\left\{
\begin{array}{lcl}
\displaystyle\left[m\Gamma(2-\gamma+\omega)\right]^{\frac{1}{2-\gamma+\omega}}\lambda^{\frac{3-\gamma+\omega}{2-\gamma+\omega}} & , & 1<\gamma-\omega<2 \\
 & & \\
\displaystyle\left[\frac{\lambda^{\gamma-\omega-1}\left(\gamma-\omega-1\right)}{A_{3}(1)}\right]^{\frac{1}{\gamma-\omega-2}}\left(\lambda-\lambda_{c}\right)^{\frac{1}{\gamma-\omega-2}} & , & 2<\gamma-\omega<3 \\
 & & \\
\displaystyle\frac{2}{\lambda\left[m\left(1+2\lambda\right)\left(\frac{\gamma-\omega-2}{\gamma-\omega-3}\right)-1\right]}\left(\lambda-\lambda_{c}\right) & , & \gamma-\omega>3
\end{array}
\right.,
\label{Thetatsis}
\end{align}
where $A_{3}(1):=\int_{0}^{\infty}\frac{\textup{d}y}{y^{\gamma-\omega-2}}\frac{e^{-y}\left(\lambda^{-1}+1+e^{-y}\right)}{\left(\lambda^{-1}+1-e^{-y}\right)^{3}}$.

\begin{figure}
\begin{center}
\includegraphics[width=243pt]{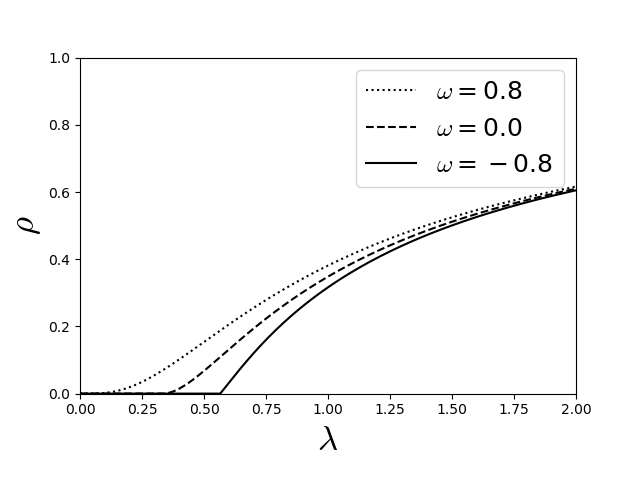}
\caption{\label{tsisprevalencexlambda} Graph $\rho\times \lambda$ for the t-SIS model with weakened connections; here, $m=1$ and $\gamma=2.5$. The curves were obtained from numerical resolution of equation \eqref{principalT}.}
\end{center}
\end{figure}

We discuss this result comparing to the ones obtained in the s-SIS model in the next section.

 %%%%%%%%%%%%%%%%%%%%%%%%%%%%%%%%%%%%%%%%%%%%%%%%%%%%%%%%%%%%%%%%%%%%%%%%%%%%%%%%%%%%%%%%%%%%%%%%%%%%
 %%%%%%%%%%%%%%%%%%%%%%%%%%%%%%%%%%%%%%%%%%%%%%%%%%%%%%%%%%%%%%%%%%%%%%%%%%%%%%%%%%%%%%%%%%%%%%%%%%%%
 %%%%%%%%%%%%%%%%%%%%%%%%%%%%%%%%%%%%%%%%%%%%%%%%%%%%%%%%%%%%%%%%%%%%%%%%%%%%%%%%%%%%%%%%%%%%%%%%%%%%

\section{Comparison of s-SIS and t-SIS models with weakened connections}
\label{comparison}

\begin{figure}
\begin{center}
\includegraphics[width=200pt]{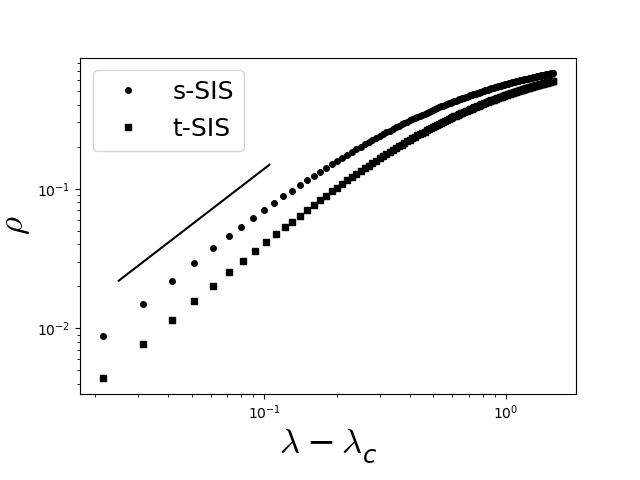}
\includegraphics[width=200pt]{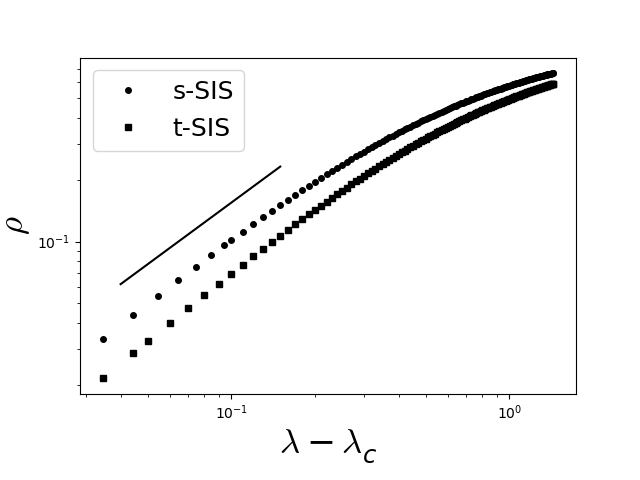}
\caption{\label{loglognontrivialmf} Graph $\rho\times (\lambda-\lambda_{c})$ in log-log scale, with $\gamma=3.25$ and $m=1$; in the left figure, $\omega=0.5$, while $\omega=0.0$ in the right graph. The lines are proportional to $\left(\lambda-\lambda_{c}\right)^{\frac{1}{\gamma-\omega-2}}$ (left figure) and $\left(\lambda-\lambda_{c}\right)^{1}$ (right figure).}
\end{center}
\end{figure}

The s-SIS and t-SIS models, as seen in the previous sections, share some common features like the infection threshold and some critical exponents. In the region $2<\gamma-\omega<3$, both models have the same nontrivial exponent $\frac{1}{\gamma-\omega-2}$, and if $\gamma-\omega>3$, they displayed the mean-field exponent $1$. One can see in figure \ref{loglognontrivialmf} that although the prevalence curve is higher in s-SIS than in the t-SIS model (as pointed out by \cite{M16}), the critical exponents are the same.

\begin{figure}
\begin{center}
\includegraphics[width=243pt]{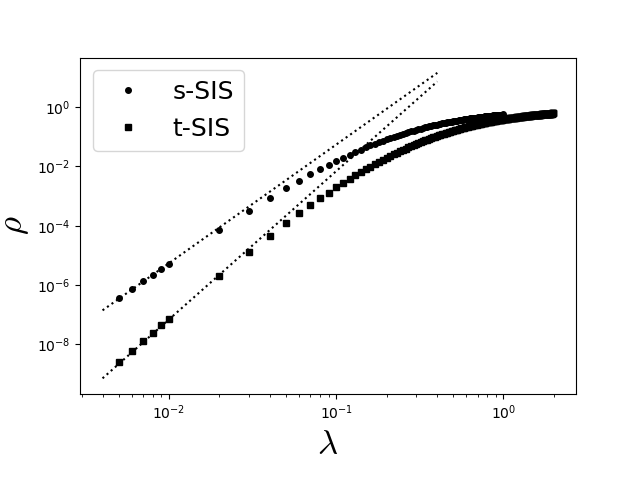}
\caption{\label{loglogsub} Graph $\rho\times\lambda$ in log-log scale, with $\gamma=2.25$, $\omega=0.5$ and $m=1$. The upper dotted line is proportional to $\lambda^{\frac{1}{2-\gamma+\omega}}$, while the lower one is proportional to $\lambda^{\frac{3-\gamma+\omega}{2-\gamma+\omega}}$; they were drawn just for visual aid.}
\end{center}
\end{figure}

The main difference between these two models, however, is observed when the infection threshold vanishes. As one can see from \eqref{Thetassis} and \eqref{Thetatsis}, the exponent with which the prevalence decays to zero differs in both cases, and this fact is supported by numerical calculations, as seen in figure \ref{loglogsub}.

\begin{figure}
\begin{center}
\includegraphics[width=243pt]{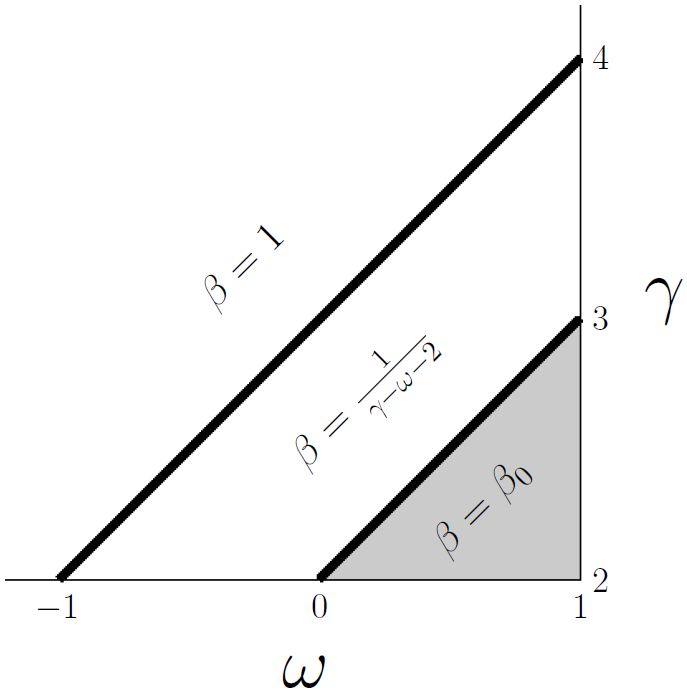}
\caption{\label{gammaomega} Phase diagram $\gamma\times\omega$. As usual, the exponent $\beta$ is defined through $\rho\sim\left(\lambda-\lambda_{c}\right)^{\beta}$. The shaded region corresponds to $\lambda_{c}=0$; otherwise, $\lambda_{c}>0$. Here, $\beta_{0}=\frac{1}{2-\gamma+\omega}$ and $\beta_{0}=\frac{3-\gamma+\omega}{2-\gamma+\omega}$ for the s-SIS and t-SIS HMF models, respectively. No algebraic decay of the order parameter is observed along the thick lines.}
\end{center}
\end{figure}

The above results can be summarized by the diagram of figure \ref{gammaomega}. We noticed that the weakening effect in connections, as defined in \eqref{qw}, can shift the critical behaviour of the prevalence. Furthermore, the weakening parameter $\omega$ and the scale-free exponent $\gamma$ are closely tied in the probability $\Theta_{\omega}$, which depends on the difference $\gamma-\omega$. This observations extends also to the prefactor of the leading term of $\Theta_{\omega}$ in both s-SIS and t-SIS models, as one can see in \eqref{Thetassis} and \eqref{Thetatsis}. However, the same property is not shared with the prevalence $\rho$, where $\gamma$ and $\omega$ become two independent parameters, as \eqref{rhossisafter} and \eqref{rhotsisafter} indicate: although all the critical exponents are functions of the difference $\gamma-\omega$ also in the prevalence, the same can not be said to the prefactors of both s-SIS and t-SIS.

 %%%%%%%%%%%%%%%%%%%%%%%%%%%%%%%%%%%%%%%%%%%%%%%%%%%%%%%%%%%%%%%%%%%%%%%%%%%%%%%%%%%%%%%%%%%%%%%%%%%%
 %%%%%%%%%%%%%%%%%%%%%%%%%%%%%%%%%%%%%%%%%%%%%%%%%%%%%%%%%%%%%%%%%%%%%%%%%%%%%%%%%%%%%%%%%%%%%%%%%%%%
 %%%%%%%%%%%%%%%%%%%%%%%%%%%%%%%%%%%%%%%%%%%%%%%%%%%%%%%%%%%%%%%%%%%%%%%%%%%%%%%%%%%%%%%%%%%%%%%%%%%%

\section{Discussion and conclusions}
\label{conclusions}

In this work, we have investigated two versions of the SIS epidemic model by a heterogeneous mean-field approach. These two models, s-SIS and t-SIS, have different transmission mechanisms, but share some properties like the location of the infection threshold and similar critical behaviour in some cases.

The main contributions of this work can be summarized as follows. Firstly, we have included and examined the influence of a weakening factor in the connections between vertices. This modification, characterised by a weakening factor $\omega$ in the linking probabilities of the underlying network of connections, may simulate a mitigation in the contact between people. We could observe that the role played by this parameter was, effectively, a redefinition of the scale-free exponent $\gamma$ to $\gamma-\omega$ for the probability $\Theta_{\omega}$ of a link reaching an infected node. This property holds for the critical exponents for the prevalence, but does not extend to its nonuniversal details like the prefactor near the infection threshold.

Secondly, we investigated both models defined on a scale-free network with degree distribution $P(k)\sim k^{-\gamma}$. While the s-SIS model has already been studied, we could characterise the decay to the critical point of the t-SIS dynamics, whose analytical critical properties seem to lack in the literature. We have shown that the prevalence decays as $\rho\sim\left(\lambda-\lambda_{c}\right)^{\beta}$ for both s-SIS nd t-SIS models, and the exponent $\beta$ is the same in some cases: it assumes a nontrivial value $\frac{1}{2-\gamma+\omega}$ for $2<\gamma-\omega<3$, and recovers the mean-field exponent $1$ for $\gamma-\omega>3$. The novelty is found when the critical point vanishes: the exponent of the prevalence in the vicinity of zero infection rate is different in s-SIS and t-SIS dynamics: in the former, the stationary infection probability decays as $\lambda^{\frac{1}{2-\gamma+\omega}}$, while in the latter, we observed $\rho\sim\lambda^{\frac{3-\gamma+\omega}{2-\gamma+\omega}}$. These results are supported by numerical methods.

 %%%%%%%%%%%%%%%%%%%%%%%%%%%%%%%%%%%%%%%%%%%%%%%%%%%%%%%%%%%%%%%%%%%%%%%%%%%%%%%%%%%%%%%%%%%%%%%%%%%%
 %%%%%%%%%%%%%%%%%%%%%%%%%%%%%%%%%%%%%%%%%%%%%%%%%%%%%%%%%%%%%%%%%%%%%%%%%%%%%%%%%%%%%%%%%%%%%%%%%%%%
 %%%%%%%%%%%%%%%%%%%%%%%%%%%%%%%%%%%%%%%%%%%%%%%%%%%%%%%%%%%%%%%%%%%%%%%%%%%%%%%%%%%%%%%%%%%%%%%%%%%%

\renewcommand{\thesection}{\Alph{section}}
\setcounter{section}{0}

\section{Appendix \thesection}
\label{apA}

In this Appendix, we examine the expansion of the integral
\begin{align}
I(\epsilon) := \int_{\epsilon}^{\infty}\frac{\textup{d}x}{x^{\alpha}\left(1+x\right)}
\label{a1}
\end{align}
for $\epsilon\sim 0$ and $\alpha>0$. In the main text, we notice that $\alpha=\gamma-\omega-1$. We consider the following cases: (i) $0<\alpha<1$, (ii) $1<\alpha<2$, (iii) $\alpha>2$ and (iv) $\alpha=1$ and $\alpha=2$.

\bigskip
\noindent
(i) Case $0<\alpha<1$. Since $I(0)<\infty$, we can write \eqref{a1} as
\begin{align}
\nonumber I(\epsilon) &= \int_{0}^{\infty}\frac{\textup{d}x}{x^{\alpha}\left(1+x\right)} - \int_{0}^{\epsilon}\frac{\textup{d}x}{x^{\alpha}\left(1+x\right)} \\
 &= B(1-\alpha,\alpha) - \int_{0}^{\epsilon}\frac{\textup{d}x}{x^{\alpha}}\left(1-x+x^{2}-x^{3}+\cdots\right),
\label{a2}
\end{align}
where $B(1-\alpha,\alpha)=\frac{\Gamma(1-\alpha)\Gamma(\alpha)}{\Gamma(1)}$ is the Beta function. Performing the integrals, we have
\begin{align}
I(\epsilon) = \Gamma(1-\alpha)\Gamma(\alpha) - \frac{\epsilon^{1-\alpha}}{1-\alpha} + \mathcal{O}(\epsilon^{2-\alpha}), \quad (0<\alpha<1).
\label{a3}
\end{align}
\bigskip
\noindent
(ii) Case $1<\alpha<2$. Here, the integral $I(0)$ diverges, but isolating the dominant term (that diverges if $\epsilon\downarrow 0$), we have
\begin{align}
I(\epsilon) = \int_{\epsilon}^{\infty}\frac{\textup{d}x}{x^{\alpha}} - \int_{\epsilon}^{\infty}\frac{\textup{d}x}{x^{\alpha-1}\left(1+x\right)},
\label{a4}
\end{align}
where the last integral in \eqref{a4} converges for $\epsilon=0$ ($1<\alpha<2$). Then, one can cast \eqref{a4} as
\begin{align}
\nonumber I(\epsilon) &= \int_{\epsilon}^{\infty}\frac{\textup{d}x}{x^{\alpha}} - \int_{0}^{\infty}\frac{\textup{d}x}{x^{\alpha-1}\left(1+x\right)} + \int_{0}^{\epsilon}\frac{\textup{d}x}{x^{\alpha-1}\left(1+x\right)} \\
\nonumber &= \frac{\epsilon^{1-\alpha}}{\alpha-1} - B(2-\alpha,\alpha-1) + \int_{0}^{\epsilon}\frac{\textup{d}x}{x^{\alpha-1}}\left(1-x+\cdots\right) \\
 &= \frac{\epsilon^{1-\alpha}}{\alpha-1} - \frac{\Gamma(2-\alpha)\Gamma(\alpha-1)}{\Gamma(1)} + \mathcal{O}(\epsilon^{2-\alpha}).
\label{a5}
\end{align}

\bigskip
\noindent
(iii) Case $\alpha>2$. Consider, initially, the case $2<\alpha<3$. Starting from \eqref{a4}, the last integral in RHS now diverges for $\epsilon\rightarrow 0$. We proceed as in the previous case by isolating the dominant term (that is responsible for the divergence); then,
\begin{align}
I(\epsilon) = \int_{\epsilon}^{\infty}\frac{\textup{d}x}{x^{\alpha}} - \int_{\epsilon}^{\infty}\frac{\textup{d}x}{x^{\alpha-1}} + \int_{\epsilon}^{\infty}\frac{\textup{d}x}{x^{\alpha-2}\left(1+x\right)}.
\label{a6}
\end{align}
The last integral is well-defined at $\epsilon=0$ for $2<\alpha<3$; therefore, writing $\int_{\epsilon}^{\infty}\frac{\textup{d}x}{x^{\alpha-2}\left(1+x\right)}=\int_{0}^{\infty}\frac{\textup{d}x}{x^{\alpha-2}\left(1+x\right)}-\int_{0}^{\epsilon}\frac{\textup{d}x}{x^{\alpha-2}\left(1+x\right)}$ leads to
\begin{align}
\nonumber I(\epsilon) &= \frac{\epsilon^{1-\alpha}}{\alpha-1} - \frac{\epsilon^{2-\alpha}}{\alpha-2} + B(3-\alpha,\alpha-2) - \int_{0}^{\epsilon}\frac{\textup{d}x}{x^{\alpha-2}\left(1+x\right)} \\
 &= \frac{\epsilon^{1-\alpha}}{\alpha-1} - \frac{\epsilon^{2-\alpha}}{\alpha-2} + \frac{\Gamma(3-\alpha)\Gamma(\alpha-2)}{\Gamma(1)} + \mathcal{O}(\epsilon^{3-\alpha})
\label{a7}
\end{align}
for $2<\alpha<3$. Now, even if $\alpha\geq 3$, the first two dominant terms are the same from \eqref{a7}; therefore,
\begin{align}
I(\epsilon) = \frac{\epsilon^{1-\alpha}}{\alpha-1} - \frac{\epsilon^{2-\alpha}}{\alpha-2} + \mathcal{O}(1)
\label{a8}
\end{align}
for $\alpha>2$.

\bigskip
\noindent
Cases $\alpha=1$ and $\alpha=2$. Here, the integrals can be performed by partial fraction decomposition. The results are
\begin{align}
I(\epsilon) = \int_{\epsilon}^{\infty}\frac{\textup{d}x}{x\left(1+x\right)} = \ln\left(1+\frac{1}{\epsilon}\right) \quad (\alpha=1)
\label{a9}
\end{align}
and
\begin{align}
I(\epsilon) = \int_{\epsilon}^{\infty}\frac{\textup{d}x}{x^{2}\left(1+x\right)} = \frac{1}{\epsilon} - \ln\left(1+\frac{1}{\epsilon}\right) \quad (\alpha=2).
\label{a10}
\end{align}

Recollecting the results above, we have
\begin{align}
I(\epsilon) =\left\{
\begin{array}{lcl}
\frac{\pi}{\sin(\alpha\pi)} - \frac{\epsilon^{1-\alpha}}{1-\alpha} + \mathcal{O}(\epsilon^{2-\alpha}) & , & 0<\alpha<1 \\
 & & \\
-\ln\epsilon + \epsilon + \mathcal{O}(\epsilon^{2}) & , & \alpha=1 \\
 & & \\
\frac{\epsilon^{1-\alpha}}{\alpha-1} - \left[\frac{\pi}{\sin(-\alpha\pi)}\right] + \mathcal{O}(\epsilon^{2-\alpha}) & , & 1<\alpha<2 \\
 & & \\
\epsilon^{-1} + \ln\epsilon + \mathcal{O}(\epsilon) & , & \alpha=2 \\
 & & \\
\frac{\epsilon^{1-\alpha}}{\alpha-1} - \frac{\epsilon^{2-\alpha}}{\alpha-2} + \mathcal{O}(1) & , & \alpha>2
\end{array}
\right.,
\label{a11}
\end{align}
where the identity $\Gamma(1-\alpha)\Gamma(\alpha)=\frac{\pi}{\sin(\alpha\pi)}$ ($\alpha\notin\mathbb{Z}$) was invoked.

 %%%%%%%%%%%%%%%%%%%%%%%%%%%%%%%%%%%%%%%%%%%%%%%%%%%%%%%%%%%%%%%%%%%%%%%%%%%%%%%%%%%%%%%%%%%%%%%%%%%%
 %%%%%%%%%%%%%%%%%%%%%%%%%%%%%%%%%%%%%%%%%%%%%%%%%%%%%%%%%%%%%%%%%%%%%%%%%%%%%%%%%%%%%%%%%%%%%%%%%%%%
 %%%%%%%%%%%%%%%%%%%%%%%%%%%%%%%%%%%%%%%%%%%%%%%%%%%%%%%%%%%%%%%%%%%%%%%%%%%%%%%%%%%%%%%%%%%%%%%%%%%%

\section{Appendix \thesection}
\label{apB}

In this Appendix, the asymptotic regime of the integral
\begin{align}
A_{0}(\varphi) := \int_{m}^{\infty}\frac{\textup{d}x}{x^{\beta}}\frac{1}{1+\lambda\left(1-\varphi^{x}\right)}
\label{b1}
\end{align}
is examined. We will omit the dependence on the parameters $\beta$ and $\lambda$ to keep the notation simpler (similar comments for the forthcoming functions). Here, the parameters obey $\beta>1$ (in the main text, $\beta=\gamma-\omega$), $\lambda>0$ and the variable $\varphi$ is restricted to $0\leq\varphi<1$. It is important to remember that $\varphi$ is strictly smaller than $1$, because we are searching for the nontrivial solution of \eqref{principalT}, as explained in section \ref{tSISw}. Introducing a change of variable $y=-x\ln\varphi$, the integral \eqref{b1} can be written as
\begin{align}
A_{0}(\varphi, \lambda) = \frac{\left(-\ln\varphi\right)^{\beta-1}}{\lambda}\int_{-m\ln\varphi}^{\infty}\frac{\textup{d}y}{y^{\beta}}\frac{1}{\lambda^{-1}+1-e^{-y}}.
\label{b2}
\end{align}

One should now to examine the behaviour of the integral
\begin{align}
A_{1}(\varphi) := \int_{-m\ln\varphi}^{\infty}\frac{\textup{d}y}{y^{\beta}}\frac{1}{\lambda^{-1}+1-e^{-y}} \quad (\beta>1)
\label{b3}
\end{align}
for $\varphi\sim 1^{-}$. The general scheme consists of integrating by parts until the exponent of the algebraic term in the integrand (in \eqref{b3}, it is $\beta$) falls in the interval $(0,1)$. When this condition is reached, the resulting integral can be replaced by its value at $\varphi=1^{-}$ without diverging, and the correction terms can be easily managed. We illustrate this procedure by dealing with the following cases: (i) $1<\beta<2$, (ii) $2<\beta<3$ and (iii) $\beta>3$.

\bigskip
\noindent
(i) Case $1<\beta<2$. One should first note that the integral in \eqref{b3} diverges for $\varphi\rightarrow 1$. By performing an integration by parts, one has
\begin{align}
A_{1}(\varphi) := \frac{1}{\beta-1}\left[\frac{\left(-m\ln\varphi\right)^{1-\beta}}{\lambda^{-1}+1-\varphi^{m}} - A_{2}(\varphi)\right],
\label{b4}
\end{align}
where
\begin{align}
A_{2}(\varphi) := \int_{-m\ln\varphi}^{\infty}\frac{\textup{d}y}{y^{\beta-1}}\frac{e^{-y}}{\left(\lambda^{-1}+1-e^{-y}\right)^{2}}.
\label{b5}
\end{align}
For $1<\beta<2$, the integral $A_{2}(\varphi)$ converges at $\varphi=1^{-}$. Then,
\begin{align}
\nonumber A_{2}(\varphi) &= A_{2}(1) - \int_{0}^{-m\ln\varphi}\frac{\textup{d}y}{y^{\beta-1}}\left[\frac{e^{-y}}{\left(\lambda^{-1}+1-e^{-y}\right)^{2}}\right] \\
\nonumber &= A_{2}(1) - \int_{0}^{-m\ln\varphi}\frac{\textup{d}y}{y^{\beta-1}}\left[\lambda^{2} + \mathcal{O}(y)\right] \\
 &= A_{2}(1) - \lambda^{2}\frac{\left(-m\ln\varphi\right)^{2-\beta}}{2-\beta} + \mathcal{O}(\ln^{3-\beta}\varphi).
\label{b6}
\end{align}
Therefore, from \eqref{b2} to \eqref{b6}, we have
\begin{align}
A_{0}(\varphi) &= \frac{1}{\beta-1}\Bigg[\frac{m^{1-\beta}}{1+\lambda\left(1-\varphi^{m}\right)} - \frac{\left(-\ln\varphi\right)^{\beta-1}}{\lambda}A_{2}(1) + \lambda\frac{m^{2-\beta}\left(-\ln\varphi\right)}{2-\beta}\Bigg] + \mathcal{O}(\ln^{2}\varphi) \quad (1<\beta<2).
\label{b7}
\end{align}

\bigskip
\noindent
(ii) Case $2<\beta<3$. The starting point is the integral \eqref{b5}. Contrary to the previous case, now the integral $A_{2}$ diverges for $\varphi\rightarrow 1^{-}$. Then, integrating it by parts leads to
\begin{align}
A_{2}(\varphi) = \frac{1}{\beta-2}\left[ \frac{\left(-m\ln\varphi\right)^{2-\beta}\varphi^{m}}{\left(\lambda^{-1}+1-\varphi^{m}\right)^{2}} - A_{3}(\varphi) \right],
\label{b8}
\end{align}
where
\begin{align}
A_{3}(\varphi) := \int_{-m\ln\varphi}^{\infty}\frac{\textup{d}y}{y^{\beta-2}}\frac{e^{-y}\left(\lambda^{-1}+1+e^{-y}\right)}{\left(\lambda^{-1}+1-e^{-y}\right)^{3}}.
\label{b9}
\end{align}
For $2<\beta<3$, $A_{3}$ converges for $\varphi\rightarrow 1^{-}$. Then,
\begin{align}
\nonumber A_{3}(\varphi) &= A_{3}(1) - \int_{0}^{-m\ln\varphi}\frac{\textup{d}y}{y^{\beta-2}}\left[\frac{e^{-y}\left(\lambda^{-1}+1+e^{-y}\right)}{\left(\lambda^{-1}+1-e^{-y}\right)^{3}}\right] \\
\nonumber &= A_{3}(1) - \int_{0}^{-m\ln\varphi}\frac{\textup{d}y}{y^{\beta-2}}\left[\lambda^{2}\left(1+2\lambda\right) + \mathcal{O}(y)\right] \\
 &= A_{3}(1) - \lambda^{2}\left(1+2\lambda\right)\frac{\left(-m\ln\varphi\right)^{3-\beta}}{3-\beta} + \mathcal{O}(\ln^{4-\beta}\varphi).
\label{b10}
\end{align}
From \eqref{b2}, \eqref{b3}, \eqref{b4}, \eqref{b8} and \eqref{b10}, one has
\begin{align}
A_{0}(\varphi) &= \frac{1}{\lambda\left(\beta-1\right)}\Bigg[ \frac{m^{1-\beta}}{\lambda^{-1}+1-\varphi^{m}} -\frac{m^{2-\beta}\left(-\ln\varphi\right)\varphi^{m}}{\left(\beta-2\right)\left(\lambda^{-1}+1-\varphi^{m}\right)^{2}} + \frac{\left(-\ln\varphi\right)^{\beta-1}}{\beta-2}A_{3}(1) + \mathcal{O}(\ln^{2}\varphi)\Bigg] \quad (2<\beta<3).
\label{b11}
\end{align}

\bigskip
\noindent
(iii) Case $\beta>3$. The starting point is the integral \eqref{b9}. Contrary to the previous case, now the integral $A_{3}$ diverges for $\varphi\rightarrow 1^{-}$. Then, integrating it by parts leads to
\begin{align}
A_{3}(\varphi) &= \frac{1}{\beta-3}\Bigg[ \frac{\left(-m\ln\varphi\right)^{3-\beta}\varphi^{m}\left(\lambda^{-1}+1+\varphi^{m}\right)}{\left(\lambda^{-1}+1-\varphi^{m}\right)^{3}} - A_{4}(\varphi) \Bigg],
\label{b12}
\end{align}
where $A_{4}(\varphi)$ is an integral that appears after the procedure above. We will not show its explicit form, since we have sufficient terms in the expansion of $A_{0}(\varphi)$ and the contribution of $A_{4}(\varphi)$ will be negligible. If $3<\beta<4$, $A_{4}$ is of order $\mathcal{O}(1)$, while $A_{4}(\varphi)=\mathcal{O}(\ln^{4-\beta}\varphi)$ if $\beta>4$; then, $A_{4}(\varphi)=\mathcal{O}(1)$ for $\beta>3$. Hence, from \eqref{b2}, \eqref{b3}, \eqref{b4}, \eqref{b8} and \eqref{b12}, one finally has
\begin{align}
\nonumber A_{0}(\varphi) &= \frac{1}{\lambda\left(\beta-1\right)}\Bigg[ \frac{m^{1-\beta}}{\lambda^{-1}+1-\varphi^{m}} -\frac{m^{2-\beta}\left(-\ln\varphi\right)\varphi^{m}}{\left(\beta-2\right)\left(\lambda^{-1}+1-\varphi^{m}\right)^{2}} + \frac{m^{3-\beta}\left(-\ln\varphi\right)^{2}\varphi^{m}\left(\lambda^{-1}+1+\varphi^{m}\right)}{\left(\beta-2\right)\left(\beta-3\right)\left(\lambda^{-1}+1-\varphi^{m}\right)^{3}} + \\
 &+ \mathcal{O}(\ln^{3}\varphi,\ln^{\beta-1}\varphi)\Bigg] \quad (\beta>3).
\label{b13}
\end{align}
Summarizing the results above, we have
\begin{widetext}
\begin{align}
A_{0}(\varphi) = \left\{
\begin{array}{lcl}
\frac{1}{\lambda\left(\beta-1\right)}\Bigg[\frac{m^{1-\beta}}{\lambda^{-1}+1-\varphi^{m}} - \left(-\ln\varphi\right)^{\beta-1}A_{2}(1) + \lambda^{2}\frac{m^{2-\beta}\left(-\ln\varphi\right)}{2-\beta} \Bigg] + \mathcal{O}(\ln^{2}\varphi) & , & 1<\beta<2 \\
\frac{1}{\lambda\left(\beta-1\right)}\Bigg[ \frac{m^{1-\beta}}{\lambda^{-1}+1-\varphi^{m}} -\frac{m^{2-\beta}\left(-\ln\varphi\right)\varphi^{m}}{\left(\beta-2\right)\left(\lambda^{-1}+1-\varphi^{m}\right)^{2}} + \frac{\left(-\ln\varphi\right)^{\beta-1}}{\beta-2}A_{3}(1) \Bigg] + \mathcal{O}(\ln^{2}\varphi) & , & 2<\beta<3 \\
\frac{1}{\lambda\left(\beta-1\right)}\Bigg[ \frac{m^{1-\beta}}{\lambda^{-1}+1-\varphi^{m}} -\frac{m^{2-\beta}\left(-\ln\varphi\right)\varphi^{m}}{\left(\beta-2\right)\left(\lambda^{-1}+1-\varphi^{m}\right)^{2}} + \frac{m^{3-\beta}\left(-\ln\varphi\right)^{2}\varphi^{m}\left(\lambda^{-1}+1+\varphi^{m}\right)}{\left(\beta-2\right)\left(\beta-3\right)\left(\lambda^{-1}+1-\varphi^{m}\right)^{3}} \Bigg] + \mathcal{O}(\ln^{3}\varphi,\ln^{\beta-1}\varphi) & , & \beta>3
\end{array}
\right..
\label{b14}
\end{align}
\end{widetext}
Translating the result \eqref{b14} into the variable $\Theta=1-\varphi$, we have

\begin{align}
A_{0}(\Theta) = \left\{
\begin{array}{lcl}
\frac{1}{\beta-1}\left[ m^{1-\beta} - \frac{A_{2}(1)}{\lambda}\Theta^{\beta-1} \right] + \mathcal{O}(\Theta) & , & 1<\beta<2 \\
\frac{1}{\beta-1}\Bigg[ m^{1-\beta} - \lambda\left(\frac{\beta-1}{\beta-2}\right)m^{2-\beta}\Theta + \frac{A_{3}(1)}{\lambda\left(\beta-2\right)}\Theta^{\beta-1} \Bigg] + \mathcal{O}(\Theta^{2}) & , & 2<\beta<3 \\
\frac{1}{\beta-1}\Bigg[ m^{1-\beta} - \lambda\left(\frac{\beta-1}{\beta-2}\right)m^{2-\beta}\Theta + \frac{\lambda m^{2-\beta}\left(\beta-1\right)}{2}\left[ \frac{m\left(1+2\lambda\right)}{\beta-3} - \frac{1}{\beta-2} \right]\Theta^{2} \Bigg] + \mathcal{O}(\Theta^{3},\Theta^{\beta-1}) & , & \beta>3
\end{array}
\right..
\label{b15}
\end{align}
when $\Theta\sim 0$. We also note that $A_{2}(1)\sim\lambda^{2}\Gamma(2-\beta)$ for $\lambda\sim 0$, which is the neighbourhood of the null infection threshold when $\beta<2$.

 %%%%%%%%%%%%%%%%%%%%%%%%%%%%%%%%%%%%%%%%%%%%%%%%%%%%%%%%%%%%%%%%%%%%%%%%%%%%%%%%%%%%%%%%%%%%%%%%%%%%
 %%%%%%%%%%%%%%%%%%%%%%%%%%%%%%%%%%%%%%%%%%%%%%%%%%%%%%%%%%%%%%%%%%%%%%%%%%%%%%%%%%%%%%%%%%%%%%%%%%%%
 %%%%%%%%%%%%%%%%%%%%%%%%%%%%%%%%%%%%%%%%%%%%%%%%%%%%%%%%%%%%%%%%%%%%%%%%%%%%%%%%%%%%%%%%%%%%%%%%%%%%

\end{document}